Journal of Risk and Financial Management

MDPI

*Article*

# Portfolio Volatility Estimation Relative to Stock Market Cross-Sectional Intrinsic Entropy

Claudiu Vinţe [1,*] and Marcel Ausloos [2,3,4]

1. Department of Economic Informatics and Cybernetics, Bucharest University of Economic Studies, 010552 Bucharest, Romania
2. School of Business, Brookfield, University of Leicester, Leicester LE2 1RQ, UK
3. Department of Statistics and Econometrics, Bucharest University of Economic Studies, 010374 Bucharest, Romania
4. GRAPES (Group of Researchers for Applications of Physics in Economy and Sociology), 483 Rue de la Belle Jardiniere, B-4031 Liege, Belgium
* Correspondence: claudiu.vinte@ie.ase.ro; Tel.: +40-751-251-119

**Abstract:** Selecting stock portfolios and assessing their relative volatility risk compared to the market as a whole, market indices, or other portfolios is of great importance to professional fund managers and individual investors alike. Our research uses the cross-sectional intrinsic entropy (*CSIE*) model to estimate the cross-sectional volatility of the stock groups that can be considered together as portfolio constituents. The *CSIE* market volatility estimate is based on daily traded prices—open, high, low, and close (OHLC)—along with the daily traded volume for symbols listed on the considered market. In our study, we benchmark portfolio volatility risks against the volatility of the entire market provided by the *CSIE* and the volatility of market indices computed using longitudinal data. This article introduces *CSIE*-based betas to characterise the relative volatility risk of the portfolio against market indices and the market as a whole. We empirically prove that, through *CSIE*-based betas, multiple sets of symbols that outperform the market indices in terms of rate of return while maintaining the same level of risk or even lower than the one exhibited by the market index can be discovered, for any given time interval. These sets of symbols can be used as constituent stock portfolios and, in connection with the perspective provided by the *CSIE* volatility estimates, to hierarchically assess their relative volatility risk within the broader context of the overall volatility of the stock market.

**Citation:** Vinţe, Claudiu, and Marcel Ausloos. 2023. Portfolio Volatility Estimation Relative to Stock Market Cross-Sectional Intrinsic Entropy. *Journal of Risk and Financial Management* 16: 114. https://doi.org/10.3390/jrfm16020114

Academic Editor: Mihaela Simionescu

Received: 17 January 2023
Revised: 6 February 2023
Accepted: 7 February 2023
Published: 11 February 2023

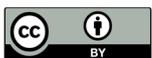

## 1. Introduction and Brief Background

Portfolio construction is as much a timing problem as a financial instrument selection problem. Deciding what to invest in may be based on the historical rate of return of the assets, which is impacted by the interest the market is showing in the particular asset and, as a corollary, the liquidity that the assets will enjoy or not during the window of time one happens to have an interest in the given asset. This window of time is intimately connected to the acquisition and possession of the portfolio.

As an introduction, it seems fair to present the broad range of proposals that have been historically investigated to tackle the problem of portfolio volatility estimation and portfolio components discovery, not only from an entropic perspective, in order to emphasise the pertinence of our aim, and its "originality".

The connection between entropy and excess market returns was investigated by Maasoumi and Racine (2002). They found significant evidence of small nonlinear unconditional serial dependence within the returns, but not conclusive evidence of superior profit opportunity when using market-switching versus buy-and-hold strategies.

Dionisio et al. (2007), questioning if the standard deviation is a good measure of risk and uncertainty, argued that entropy could present some advantages as a measure of uncertainty and simultaneously verify some basic assumptions of the portfolio management theory, namely the effect of diversification.

Liu and Chen (2012) applied type-2 fuzzy theory to the portfolio selection problem, in which the security returns are unknown and are characterised by type-2 fuzzy variables. Not having the expectation and entropy of type-2 fuzzy variable well defined, they opted to reduce first the type-2 fuzzy variable and then propose a mean-entropy model with reduced variables to be applied to the portfolio selection problem. The intention was to transform the established mean-entropy model of reduced variables into equivalent parametric programming.

Ausloos (2000) argued that in classical thermodynamics, the entropy is necessarily coupled to a temperature and that the temperature is known to mimic the inverse of a relaxation time. Similarly, he proposes that it is reasonable to assume that financial market actors may consider relaxation times on the market and that the relaxation times differ in reality, reflecting different perspectives that actors have regarding the market evolution. Nevertheless, the interaction which will be considered will be a self-interaction of the information asymmetry level of the share price (stock value), which itself is in some sense due to different appreciation conditions of the market by actors (Dhesi and Ausloos 2016).

To study portfolio diversity, Song and Chan (2020) proposed an adaptive entropy model, which incorporates entropy measurement and adaptability into the conventional Markowitz mean-variance model. Mercurio et al. (2020) introduced a new family of portfolio optimization problems called return-entropy portfolio optimization (REPO) that simplifies the computation of portfolio entropy using a combinatorial approach.

Novais et al. (2022) proposed a portfolio-optimisation model that uses entropy and mutual information as risk measurements instead of variance and covariance. They experimented by comparing models that rely on mean-variance with counterparts based on mean-entropy using a stochastic entropy estimation. Their results showed that when increasing return constraints on portfolio optimisation, the mean-entropy models were more stable overall, exhibiting dampened responses in cumulative returns and Sharpe ratio in comparison to mean-variance methods.

According to Capital Asset Pricing Model (CAPM), the only risk that investors should be compensated for is the risk that cannot be diversified away (Sharpe 1964; Ross 1976). Only systematic risk will command a risk premium. CAPM is calculated according to the following formula:

$$R_x = R_{rf} + [\beta_x \cdot (R_m - R_{rf})], \qquad (1)$$

where:

$R_x$—expected return on security $x$
$R_{rf}$—risk-free rate
$\beta_x$—beta of the security $x$
$R_m$—the expected return of the market.

The difference $(R_m - R_{rf})$ represents the risk premium. Since the risk-free rate $R_{rf}$ is pegged and readily available for a considered time interval, the expected return of the market $R_m$ is an estimation that may differ depending on the model used. There are also considerations related to the computation of returns on a stock mix, when short sales are permitted, or when short sales are not permitted (Lintner 1965), and then the risk-free rate $R_{rf}$ can play a relevant decisional role.

A market index, such as the S&P500, is not the entire market. The total market includes thousands of other traded stocks and, in a broader sense, bonds, real estate, commodities, options, and many other assets of all sorts, including one of the most important assets any of us has: the human capital built up by education, work, and life experience (Black and Litterman 1992; Malkiel [1975] 2020). There is also evidence that, when estimating the volatility of the stock market, a more inclusive instrument, in terms of the number of symbols taken into account for what constitutes the stock market, provides safer grounds for non-manipulative interpretations (Bhowmik and Wang 2020). For example, Saha et al. (2019) found that the movements in the daily levels of the VIX index are explained by market fundamentals and not by manipulation. They show that the VIX closing values and VIX futures settlement prices from 2008 are consistent with normal market forces and are not artificial.

Moreira and Muir (2017) proposed a volatility-managed portfolio strategy consisting of constructing portfolios that adjust monthly returns by the inverse of their previous month's realised variance, thus decreasing risk exposure when variance was recently high and vice versa. They documented that this trading strategy earns large alphas across a wide range of asset pricing factors, suggesting that investors can benefit from volatility timing.

Volatility estimates of a portfolio of stocks need to be considered contextually. Volatility estimated through standard deviation or variance of asset prices over time may offer an indication of the dispersion of prices around the mean calculated for a given time interval but is the performance relative to other assets on the market, or against the market as a whole, that matters more when it comes to portfolio selection. For example, Carr and Wu (2009) used a large option data set to synthesise variance swap rates and investigate the historical behaviour of variance risk premiums on five stock indexes and 35 individual stocks. Cejnek and Mair (2021) implemented timing regressions and related returns of a volatility-managed portfolio to discount rate, cash flow, and expected volatility, providing evidence that volatility management outperforms by levering up good times without increasing downside exposure to fundamental risk drivers.

Fama and French (1992) divided all traded stocks into deciles according to their beta measures over the 1963–1990 period. They concluded that there is no relationship between beta and return. Additionally, small firms tended to outperform larger firms with the same beta levels. Therefore, size is a risk factor that deserves to be compensated for with additional return. Castellano and Cerqueti (2014) analysed a mean-variance optimal portfolio selection problem in the presence of risky assets characterized by low-frequency trading and therefore low liquidity. These attributes most often describe small market capitalization companies, which are not well known by investors and whose stocks, consequently, are not traded regularly and/or not in significant volumes. On the other hand, the Fama and French three-factor model (Fama and French 1992, 1993), which takes into account the beta relative to the market index, the capitalization of the company (size), and the market price versus book value as a ratio, shows that the smaller firms are relatively risky. There is evidence (Fama and French 1995, 1996) that returns are higher for stocks with lower price-to-book ratios and smaller sizes (market capitalization).

Previous work shows that average returns on common stocks are related to firm characteristics such as size, earnings/price, cash flow/price, book-to-market equity, past sales growth, long-term past return, and short-term past returns. Additional factors to the Fama–French three-factor model (Malkiel [1975] 2020):

(a) A momentum factor to capture the tendency for rising or falling stocks to continue moving in the same direction.
(b) A liquidity factor to reflect that investors need to be paid a return premium as an incentive to hold illiquid assets.
(c) Quality of the company, as measured by such indicators as the stability of its earnings, sales growth, and its low amount of debt.

Furthermore, stock returns can be sensitive to general market swings, changes in interest and inflation rates, changes in national income, exchange rates, and other economic factors. Investigating the aggregate volatility risk factor, Barinov (2012) proposes the hypothesis that small growth firms and equity issuers are used by portfolio managers to hedge against aggregate volatility risk.

Markowitz (1959) ideas laid down in the seminal monograph on portfolio selection sparked an entire wave of emulation in academia and among practitioners, being continuously perfected in concrete implementations, re-evaluated, and extended. Wang and Xia (2002) discussed the Markowitz model and its modifications, as well as the related models based on different criteria for risk and return, but which share the same feature as the Markowitz model, namely that there is an underlying probability distribution for changes in the stock market. They considered models in which a decision does not rely on probability distributions on stock movement, though such information may still be used.

There have been attempts to estimate the volatility of the portfolio without using an estimation of a volatility matrix (the volatilities of the individual assets in the portfolio and their correlations), although the approach estimates stochastic volatility and its volatility (Alghalith 2016).

Depending on how one measures the market, different beta measures can be obtained. Searching for low beta stocks with returns as attractive as for the market as a whole but with much less risk. Or collect high-return stocks with beta on par with the market index. Traditional betas refer to the index of the stock market, as broad as it can be, and the beta of the market is defined as having a value of 1. In the case of *CSIE* market volatility estimates, a market index such as S&P500 can have a beta against the entire market between 0.1 and 0.5, which is significantly lower than 1, which would be the beta of the entire stick market. Malkiel ([1975] 2020) refers to smart betas as indicators that are intended to identify the possibility of gaining excess returns (greater than the market) by using a variety of relatively passive rules-based investment strategies that involve no more risk than would be assumed by investing in a low-cost total stock market index fund.

Betas for individual stocks are not stable over time and are very sensitive to the market proxy against which they are measured. Tracking them against the volatility of market indices can lead to different results than using the *CSIE* volatility estimates of the entire market. González-Urteaga and Rubio (2016) investigated the determinants of the cross-sectional variation of the average volatility risk premia for a representative set of portfolios sorted by volatility risk premium beta, explaining why the volatility risk premia are different across assets.

Price and quantity have been the two fundamental components of any human trade activity since the beginning of time. One buys or sells a certain quantity of a given good at a certain price, based on the credence that that is the right deal under the given market conditions. One does not enter the trade if one considers the price to be unjustified and the history does not record anything. Alternatively, one enters the deal at a certain price level and for a certain quantity because one believes that that is the adequate quantity that one would be willing to trade at that price level. In other words, if one is not entirely convinced or satisfied with the price, then the traded quantity reflects the level of trust in the considered price level.

This paper contributes to portfolio volatility estimation with an additional quantitative instrument to assist portfolio selection, based on asset volatility relative to market indices and the volatility of the market as a whole. Our study on intrinsic entropy does not necessarily aim to identify a sole means to assist portfolio selection, but rather

(a) Make use of a comprehensive cross-sectional volatility estimator, constructed taking into account all the symbols listed and traded on a given market;
(b) Identify a subset (portfolio) of symbols built based on the rate of returns and the betas relative to the volatility of the market as a whole for various time frames and intervals of historical data.

To our best knowledge, cross-sectional intrinsic entropy (*CSIE*) is the only cross-sectional volatility estimator that:

- takes into account all the listed and traded symbols of a given market;
- includes in the model not only the daily OHLC prices but also the traded volume.

The intrinsic entropy (*IE*) volatility estimator possesses two peculiar features, compared to the variance-based volatility estimators [1]:

1. Takes into account the traded volume, in addition to the price data, bringing in additional insight regarding the market inclination.
2. It is a signed volatility estimator:
    (a) high positive values of *IE* are associated with a preponderant market buy;
    (b) while high negative values of *IE* are associated with a preponderantly market sale.

Since market indices started to be traded on the exchanges as regular securities, they became one the most, if not the most, sought-after assets in portfolios by individual and institutional investors alike, investment funds, pension funds, etc. The attractiveness of stock market indices is rightly justified due to their relatively broad base of constituents and, corroborated with this, a historically proven lower exposure to risk compared to the market as a whole (Vințe and Ausloos 2022). It is as if portfolio selection is already solved by owning a single asset that offers exposure to many stocks. Although there is still a significant drawback to not owning the actual stocks by not benefiting from the dividends the issuing companies may pay annually, but the lower risk associated with the market indices can be a tractive enough compensation for many investors. Additionally, diversification is always desirable, since not a single asset, not even an exchange-traded stock market index, can offer full coverage concerning market volatility.

In such a framework, the research questions of our study are the following.

i. For any given interval of time, can at least two symbols, traded on the market, be identified that have a combined risk equal to or lower than that provided by the volatility estimates of the market index, and with a higher rate of return?
ii. If multiple symbols satisfy these constraints, can we algorithmically discover all of them?

The remainder of the article is organised as follows. Section 2, Materials and Methods, presents the portfolio volatility estimation based on the cross-sectional intrinsic entropy (*CSIE*) as the volatility estimator of a set of stocks and for the stock market as a whole, along with the intrinsic entropy (*IE*) as the volatility estimator of market indices based on longitudinal data. The methodology for computing both *CSIE* and *IE* is intimately related to the format in which the market data are available, how it is preprocessed, and structurally reorganised to allow efficient computation. Therefore, the input data and the way they are organised are presented in this Section 2, together with the algorithm for the calculation of the conditional betas. Section 3, Results, introduces the results obtained that will contour the premises in view of Section 4, Discussions, concerning the traits of the stocks that exhibit lower risk than the market index and, at the same time, move in the same direction as the index, having a positive conditional beta relative to the entire stock market. Here, we also discuss the limitations of our study and the delineation concerning future research. Section 5, Conclusions, summarises the outcome of our present investigation.

## 2. Materials and Methods

According to Markowitz (1952, 1959), calculating the volatility estimate of a given portfolio *S* of *m* assets $\{x_1, x_2, \ldots, x_m\}$ takes into account the weight of each constituent in the overall value of the portfolio and the covariances between any pair of assets. The volatility estimate is provided by the rate-of-return variance of the portfolio constituents over a given time frame, say, *n* days. It is worth noting that, while the mean-variance formulation by Markowitz offers the basis for modern portfolio selection analysis in a single period, an analytical optimal solution to the mean-variance formulation in multiperiod portfolio selection has been investigated as well (Li and Ng 2001).

$$\sigma_S^2 = w^T \cdot Cov(S) \cdot w, \tag{2}$$

where $w$ is the vector of weights or how much of the total value of the portfolio is allocated to each asset,

$$w = \begin{bmatrix} w_1 \\ w_2 \\ \vdots \\ w_m \end{bmatrix}. \tag{3}$$

If we notate with $p_{ij}$ the price of the assets $x_j, j \in [1, m]$ on day $i \in [1, n]$, then the matrix of prices for all the assets considered in the portfolio in the interval of $n$ days is the following.

$$P = \begin{bmatrix} p_{11} & p_{12} & \cdots & p_{1m} \\ p_{21} & p_{22} & \cdots & p_{2m} \\ \vdots & \vdots & \ddots & \vdots \\ p_{n1} & p_{n2} & \cdots & p_{nm} \end{bmatrix}, \tag{4}$$

The price $p_{ij}$ of asset $x_j, j \in [1, m]$ on day $i \in [1, n]$ is usually considered as being the closing price of the day. The vector of price averages in the interval is considered for all components of the portfolio.

$$\mu = [\mu_1 \; \mu_2 \cdots \mu_m], \text{ where } \mu_j = \frac{1}{n}\sum_{i=1}^{n} p_{ij}, \; j \in [1, m] \tag{5}$$

Then the covariance matrix $Cov(S)$ of the portfolio, $S$ is calculated as follows.

$$Cov(S) = \begin{bmatrix} Cov(x_1, x_1) & Cov(x_1, x_2) & \cdots & Cov(x_1, x_m) \\ Cov(x_2, x_1) & Cov(x_2, x_2) & \cdots & Cov(x_2, x_m) \\ \vdots & \vdots & \ddots & \vdots \\ Cov(x_m, x_1) & Cov(x_m, x_2) & \cdots & Cov(x_m, x_m) \end{bmatrix}, \text{ where covariance is} \tag{6}$$

$$Cov(x_k, x_l) = \frac{\sum_{i=1}^{n}(p_{ik}-\mu_k)(p_{il}-\mu_l)}{n-1}, \text{ for } k \neq l; \; k, l \in [1, m], \text{ and variance} \tag{7}$$

$$Cov(x_k, x_l) = Var(x_k, x_k) = \frac{\sum_{i=1}^{n}(p_{ik}-\mu_k)^2}{n-1}, \text{ for } k = l; \; k, l \in [1, m]. \tag{8}$$

If we calculate the differences between daily prices and the interval average for each asset, matrix-wise, we obtain the following.

$$(P - \mu) = \begin{bmatrix} (p_{11} - \mu_1) & (p_{12} - \mu_2) & \cdots & (p_{1m} - \mu_m) \\ (p_{21} - \mu_1) & (p_{22} - \mu_2) & \cdots & (p_{2m} - \mu_m) \\ \vdots & \vdots & \ddots & \vdots \\ (p_{n1} - \mu_1) & (p_{n2} - \mu_2) & \cdots & (p_{nm} - \mu_m) \end{bmatrix}, \text{ and its transpose} \tag{9}$$

$$(P - \mu)^T = \begin{bmatrix} (p_{11} - \mu_1) & (p_{21} - \mu_1) & \cdots & (p_{n1} - \mu_1) \\ (p_{12} - \mu_2) & (p_{22} - \mu_2) & \cdots & (p_{n2} - \mu_2) \\ \vdots & \vdots & \ddots & \vdots \\ (p_{1m} - \mu_m) & (p_{2m} - \mu_m) & \cdots & (p_{nm} - \mu_m) \end{bmatrix}. \tag{10}$$

The constituent covariances and volatility of the portfolio become:

$$Cov(S) = \frac{1}{n-1}[(P - \mu)^T \cdot (P - \mu)], \text{ making the substitution in (2),} \tag{11}$$

$$\sigma_S^2 = w^T \cdot \frac{1}{n-1}[(P - \mu)^T \cdot (P - \mu)] \cdot w, \tag{12}$$

where $n$ is the number of days in the considered time interval.

In the context of the cross-sectional intrinsic entropy model (*CSIE*), we consider end-of-day (EOD) data containing daily open, high, low and close (OHLC) prices along with the traded quantity (volume) of each marked listed symbol that may be selected in a set as a portfolio constituent.

Historical EOD data are sourced from https://www.eoddata.com/ 5 February 2023 and consist of a daily file containing OHLC prices and the traded volume for each listed stock on the market and traded in the given day. The collections of over 5500 files for each of the markets considered in the present study, the NYSE and the NASDAQ, are processed in such a way as to obtain a multidimensional array in the memory for allowing access longitudinally to the time series, and in cross-section for daily EOD data of the entire market. Therefore, making use of historical daily OHLC prices and volume for a period of more than 21 years, from 1 January 2001 to 28 October 2022, the data are organised in a multidimensional array, having as entry point a matrix $X$ of over 5647 rows, as the number of days of daily data, and 3321 columns, listed symbols, as of 28 October 2022, for the NYSE. Correspondingly, the matrix $X$ for the NASDAQ has 5643 rows, as the number of days of daily data, and 4937 columns, listed symbols, as of 28 October 2022.

$$X = \begin{bmatrix} x_{11} & x_{12} & \cdots & x_{1m} \\ x_{21} & x_{22} & \cdots & x_{2m} \\ \vdots & \vdots & \ddots & \vdots \\ x_{n1} & x_{n2} & \cdots & x_{nm} \end{bmatrix}, \tag{13}$$

For each symbol $j$, $j \in [1, m]$ listed and traded on the day $i$, $i \in [1, n]$ we have available a 5-tuple $x_{ij}$ of values that provide a daily informational depth.

$$x_{ij} = (x_{ij}^O, x_{ij}^H, x_{ij}^L, x_{ij}^C, x_{ij}^V), \text{ for } i \in [1, n], j \in [1, m], \tag{14}$$

where:

- $x_{ij}^O$—the open price (O) of symbol $j$ on day $i$;
- $x_{ij}^H$—the high (H) price of the symbol $j$ on day $i$;
- $x_{ij}^L$—the low (L) price of symbol $j$ on day $i$;
- $x_{ij}^C$—the close price (C) of the symbol $j$ on day $i$;
- $x_{ij}^V$—the traded volume (V) of symbol $j$ on day $i$.

The values $\{m_1, m_2, m_3, \ldots, m_i, \ldots, m_n\}$ are the number of the symbols listed and traded on the market on the corresponding day $i \in [1, n]$ and $m = \max(m_i), \forall i \in [1, n]$.

We point out that since the number of listed symbols on both the NYSE and the NASDAQ markets has changed over time, generally exhibiting an ascending trend, the matrix X which contains all the listed symbols on a given market is a fairly sparse matrix. The matrix X sparsity can exceed 50%, to provide a rough magnitude level, in the context in which matrix X has over 18.75 mils. cells for the NYSE market and more than 27.85 mils. cells for the NASDAQ. Additionally, each cell $x_{ij}$ in matrix $X$ stores a tuple of 5 values, see relation (14). It is worth noting that research motivated by arbitrage pricing theory in finance has been conducted to reduce dimensionality and to estimate the covariance matrix through a multifactor model (Fan et al. 2008, 2016; Fan and Kim 2018) or to estimate the large integrated volatility matrix without using covolatilities of illiquid assets (Fan and Kim 2019).

The daily total traded value, considered at the end of each trading day $i$, is given by the following relation:

$$\lambda_i = \sum_{j=1}^{m_i} x_{ij}^C x_{ij}^V, \qquad \text{for } i \in [1, n] \tag{15}$$

Therefore, the daily ratio of individual symbols in the overall traded value $S_i$ is defined by:

$$\psi_{ij} = \frac{x_{ij}^C x_{ij}^V}{\lambda_i}, \qquad \text{for } i \in [1, n], j \in [1, m_i]. \tag{16}$$

where $m_i$ is the number of the symbols listed and traded on the market on the corresponding day $i$. If we notate $\lambda_{ij} = x_{ij}^C x_{ij}^V$, then

$$\psi_{ij} = \frac{\lambda_{ij}}{\lambda_i}, \qquad \text{for } i \in [1, n], j \in [1, m_i]. \tag{17}$$

Such ratios $\psi_{ij}$ denote the portion of the traded value $\lambda_{ij}$ corresponding to the symbol $j$ on day $i$ in the overall value traded on day $i$, or the total amount of money $\lambda_i$ exchanged on the market for the day $i \in [1, n]$.

With the above notation, the cross-sectional intrinsic entropy (*CSIE*) (Vințe and Ausloos 2022) of a set of symbols on a given day is:

$$H_i = (1 - f_i)H_i^{OC} + f_i H_i^{OLHC}, \text{ with } i \in [1, n]. \tag{18}$$

The components $H_i^{OC}$ and $H_i^{OLHC}$ are defined as follows:

$$H_i^{OC} = -\sum_{j=1}^{m_i} \left( \frac{x_{ij}^C}{x_{ij}^O} - 1 \right) \psi_{ij} \ln \psi_{ij} \tag{19}$$

$$H_i^{OLHC} = -\sum_{j=1}^{m_i} \left[ \left( \frac{x_{ij}^H}{x_{ij}^O} - 1 \right)\left( \frac{x_{ij}^H}{x_{ij}^C} - 1 \right) + \left( \frac{x_{ij}^L}{x_{ij}^O} - 1 \right)\left( \frac{x_{ij}^L}{x_{ij}^C} - 1 \right) \right] \psi_{ij} \ln \psi_{ij} \tag{20}$$

$$f_i = \frac{\alpha - 1}{\alpha + \frac{m_i + 1}{m_i - 1}} \tag{21}$$

for $m_i \in \{m_1, m_2, m_3, \ldots, m_n\}, \forall i \in [1, n]$.

The value of $f_i$ from Equation (21) is consistent with the determination by Yang and Zhang (2000). In their influential paper on drift-independent volatility estimation using OHLC prices, they searched for an equivalent value of $f_i$, see Equation (18), for which the variance of the volatility estimator reaches the minimum. Based on the work of Rogers and Satchell (1991) and Rogers et al. (1994), who showed that $\alpha \leq 2$ by using the triangle inequality, Yang and Zhang calculated that $\alpha \leq 1.5$ for all drifts. To optimize their volatility estimator for situations exhibiting a small drift, Yang and Zhang suggested setting $\alpha = 1.34$ in practice. Since the significance of the terms $H_i^{OC}$ and $H_i^{OLHC}$ is similar to that of $V_{OC}$ and $V_{RS}$ from the Yang–Zhang volatility estimator, we followed the same rationale for using $\alpha = 1.34$ to calculate the weight $f_i$.

If we select a portfolio $S$ of symbols from the entire market $X$ and hold it for a given time span, that means that $S$ is a subset of $X$ in terms of the number of symbols and the number of days for which the volatility of such portfolio can be estimated. *CSIE* provides a daily volatility estimate for the entire market. To estimate the volatility of the market for a $t$-day interval, we calculate the moving averages of the *CSIE* for appropriate windows of $w$-day.

The contract for difference (CFD) that is offered by most online brokers to retail customers, as a means to buy and sell stocks, excludes from the stock return equation the stock dividend since the buyer of such a contract does not own the stock bearing the dividend.

Therefore, we define $V_{CSIE,t,w}^{market}$ as being the volatility estimates of the entire market based on cross-sectional intrinsic entropy (*CSIE*) and computed for $t$-day time intervals based on rolling windows of $w$-day.

Similarly, $V_{CSIE,t,w}^{S}$ is the volatility estimate of portfolio $S$, based on the *CSIE* of the portfolio, calculated for the $t$-day time interval and windows of moving averages $w$-day.

The market index volatility estimates $V_{IE,t,w}^{index}$ is computed using intrinsic entropy (*IE*) based on time series data, using the same rolling windows of $w$-day, within the time interval of $t$-day (Vințe et al. 2021).

Additionally, the volatility estimates of an individual symbol $V_{IE,t,w}^{x_j}$ is computed based on IE, using rolling windows of the same $w$-day, within the $t$-day time interval.

With these notations, we introduce the following betas.

$$\beta_{t,w}^{index} = \frac{Cov(V_{IE,t,w}^{index}, V_{CSIE,t,w}^{market})}{Var(V_{CSIE,t,w}^{market})}; \text{ beta of the market index, relative to the entire market volatility.} \quad (22)$$

$$\beta_{t,w}^{x_j} = \frac{Cov(V_{IE,t,w}^{x_j}, V_{CSIE,t,w}^{market})}{Var(V_{CSIE,t,w}^{market})}, \text{ for } j \in [1, m]; \text{ the beta of the stock } x_j, \text{ relative to the entire market volatility. The values } \{m_1, m_2, m_3, \ldots, m_i, \ldots, m_t\} \text{ are the number of the symbols listed and traded on the market on the corresponding day } i, \text{ and } m = \max(m_i), \forall i \in t\text{-day time interval.} \quad (23)$$

$$\beta_{t,w}^{S} = \frac{Cov(V_{CSIE,t,w}^{S}, V_{CSIE,t,w}^{market})}{Var(V_{CSIE,t,w}^{market})}; \text{ beta of the portfolio } S, \text{ relative to the entire market volatility.} \quad (24)$$

Algorithm 1 for computing the betas and selecting the symbols according to the dynamically imposed criteria is described as follows.

---

**Algorithm 1:** Compute the conditional betas and select the symbols according to the dynamically imposed criteria.

---
1:    Initialize the time interval for both *CSIE* and *IE* ← *t*-day
2:    Initialize the *CSIE* moving averages ← *w*-day
3:    Initialize the *IE* rolling windows ← *w*-day
4:    Initialize an empty portfolio of stocks $S \leftarrow \{\}$
5:    Compute the return rate of the index $R_{index}$ in *t*-day interval
6:    Compute *CSIE* market volatility estimates $V_{CSIE,t,w}^{market}$
7:    Compute market index *IE* volatility estimates $V_{IE,t,w}^{index}$
8:    Compute index $\beta_{t,w}^{index} = \frac{Cov(V_{IE,t,w}^{index}, V_{CSIE,t,w}^{market})}{Var(V_{CSIE,t,w}^{market})}$
9:    **for each** stock $x_j$ traded on the market in the *t*-day interval **do**
10:      Compute stock return rate $R_{x_j}$ in *t*-day interval
11:      Compute stock *IE* volatility estimates $V_{IE,t,w}^{x_j}$
12:      Compute $\beta_{t,w}^{x_j} = \frac{Cov(V_{IE,t,w}^{x_j}, V_{CSIE,t,w}^{market})}{Var(V_{CSIE,t,w}^{market})}$
13:      **if** ($\beta_{t,w}^{x_j} \leq \beta_{t,w}^{index}$) **and** ($R_{x_j} \geq R_{index}$) **then**
14:        Add the stock to the portfolio $S \xleftarrow{+} \{x_j\}$
15:      **end if**
16:    **end for**
17:    Compute portfolio $V_{CSIE,t,w}^{S}$
18:    Compute portfolio $\beta_{t,w}^{S} = \frac{Cov(V_{CSIE,t,w}^{S}, V_{CSIE,t,w}^{market})}{Var(V_{CSIE,t,w}^{market})}$
19:    **return** $\beta_{t,w}^{S}$ of the discovered portfolio

---

Portfolio *S* will be populated with symbols that, in the *t*-day given time interval, have a lower or equal risk as the market index and, at the same time, a rate of return equal to or higher than the one realized by the index.

## 3. Results

To answer the posed research questions, we study various time intervals, in particular tumultuous periods, characterised by intense market volatility and downturns. Since the aim of our current study is not to optimise portfolio allocation, the weights of all constituents of the portfolio are considered equal. The strategy of buying and holding for a medium to long period is consistent with the type of trading in the market indices.

We first present the results obtained for three intervals of time with different time spans as follows.

- 125-day trading interval, from 1 March 2022 to 26 August 2022;

- 250-day trading interval, from 5 October 2020 to 20 December 2021;
- 950-day trading interval, from 2 April 2018 to 26 August 2022.

The intention is to investigate the most recent developments in the stock market, cover the downturn that started in the spring of 2022, including the entire period of lockdowns and uncertainties in the labour market from the fall of 2020 and the whole year 2021, along with a broader perspective provided the period of the last 4 years which covers the SARS-CoV-2 pandemic. Based on the preliminary observations drawn from these three intervals of time, we proceed to study the market on an annual basis from 2001 to 2021. We comment that even the year-base study is not designed to be an exhaustive one but rather to showcase a series of evenly divided time intervals that can be easily followed and associated with events that marked the financial-economic evolution in the overall time span. For concrete forecasting purposes, exhaustive combinations of *t*-day intervals and moving averages of the *CSIE* for various windows of *w*-day should be considered further. To not clutter the graphical representation excessively, we limit the number of stock symbols to the least risky 15 symbols, if there are more than 15 companies whose stocks satisfy the constraints.

Figure 1 shows the 15 least risky stock symbols discovered in an interval of 125 trading days, from 1 March 2022 to 26 August 2022, that have a return rate higher than the NYSE S&P500 index (the vertical line for the return rate −5.77%) and a beta, relative to the entire NYSE market, lower than the one exhibited by the S&P500 index (the horizontal line for beta 0.0642). In total, 314 stocks were identified that satisfy the constraints and have a positive beta. A rolling window of 10 days has been used. In other words, a set of symbols can be identified, more than one, that have a higher rate of return than the S&P500 market index at a maximum level of risk provided as a threshold represented here by the beta of the index relative to the entire NYSE market. We point out that while the index was down by 5.77% in the considered time interval, most of the symbols identified in the set based on the imposed restrictions were concentrated around a positive rate of return of well above 7.5%.

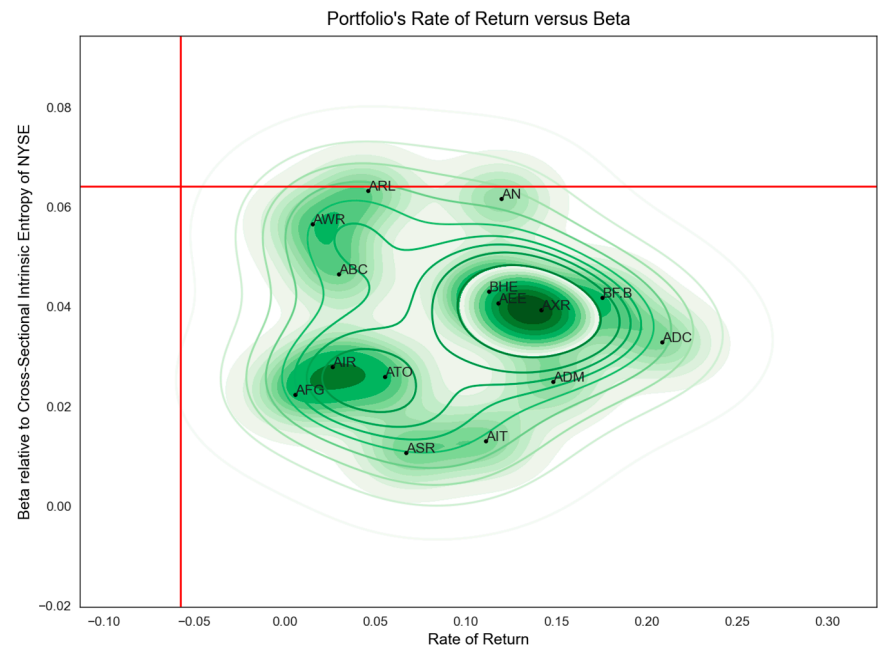

**Figure 1.** A set of 15 NYSE symbols that outperformed the S&P500 index in terms of rate of return without being exposed to higher risk (125-day trading interval, from 1 March 2022 to 26 August 2022).

Figure 2 shows the 15 least risky stock symbols discovered in an interval of 250 trading days, from 5 October 2020 to 20 December 2021, to have a rate of return higher than the NYSE S&P500 index (the vertical line for the return rate of 23.79%) and a beta, relative to the entire NYSE market, lower than the one exhibited by the S&P500 index (the horizontal line for beta 0.0175). In total, 55 stocks were identified that meet the constraints and have a positive beta. A rolling window of 20 days has been used. We point out that while the index was up by 23.79% in the considered time interval, most of the symbols identified in the set based on the imposed restrictions were concentrated around a positive rate of return of well above 40%.

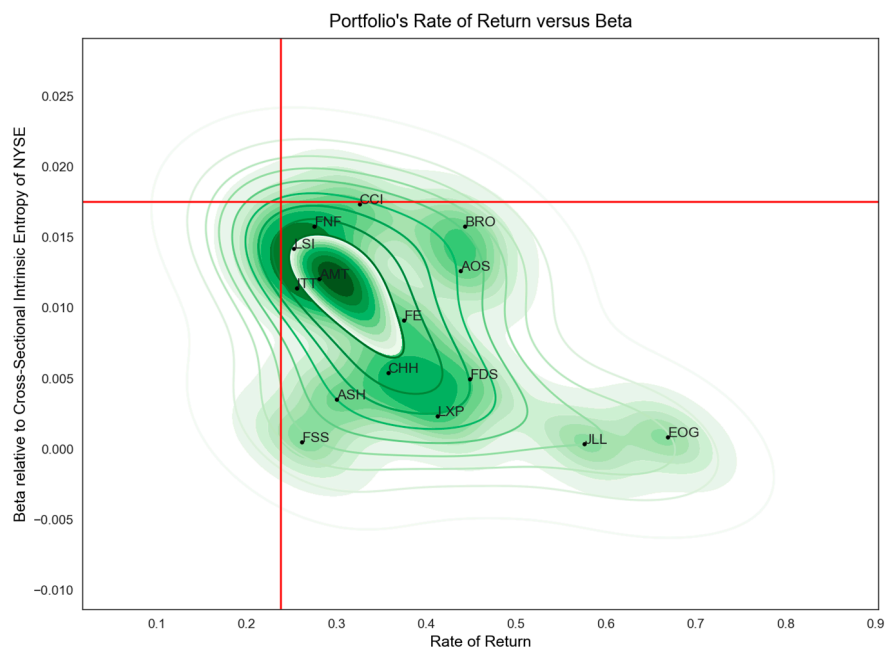

**Figure 2.** A set of 15 NYSE symbols that outperformed the S&P500 index in terms of rate of return without being exposed to higher risk (250-day trading interval, from 5 October 2020 to 20 December 2021).

Figure 3 shows the 15 least risky stock symbols discovered in an interval of 950 trading days, from 2 April 2018 to 26 August 2022, to have a rate of return higher than the NYSE S&P500 index (the vertical line for the return rate of 48.29%) and a beta, relative to the entire NYSE market, lower than the one exhibited by the S&P500 index (the horizontal line for beta 0.0369), lower than the one exhibited by the S&P500 index (the horizontal line for beta 0.0369). In total, 114 stocks were identified that meet the constraints and have a positive beta. A rolling window of 20 days has been used. It is worth noting that while the index was up by 48.29% in the considered time interval, most of the symbols identified in the set based on the imposed restrictions were concentrated around a positive rate of return of well above 75%.

Figure 4 shows the 15 least risky stock symbols discovered in an interval of 125 trading days, from 1 March 2022 to 26 August 2022, to have a rate of return higher than the NASDAQ Composite index (−10.28%) and a beta, relative to the entire NASDAQ market, lower than the one exhibited by the NASDAQ Composite index (0.0894). In total, 481 stocks were identified that meet the constraints and have a positive beta. A rolling window of 10 days has been used. In other words, a set of symbols can be identified, more than one, that have a higher rate of return than the S&P500 market index at a maximum level of risk provided as a threshold represented here by the beta of the index relative to the entire NYSE market. We point out that, while the index was down by 10.28% in the considered time interval, most of the symbols identified in the set based on the imposed restrictions were concentrated around a positive rate of return of well above 5.5%.

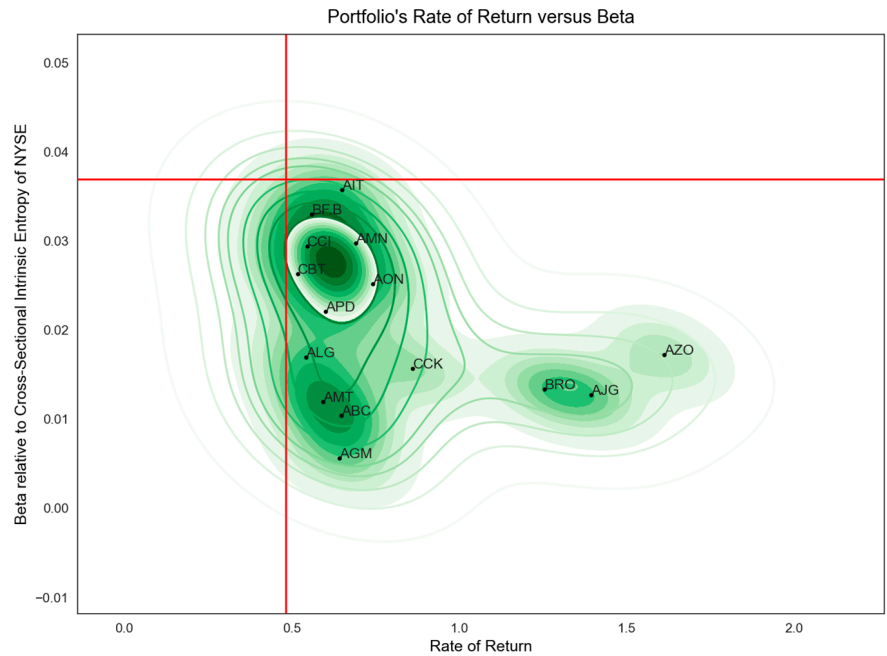

**Figure 3.** A set of 15 NYSE symbols that outperformed the S&P500 index in terms of rate of return without being exposed to higher risk (950-day trading interval, from 2 April 2018 to 26 August 2022).

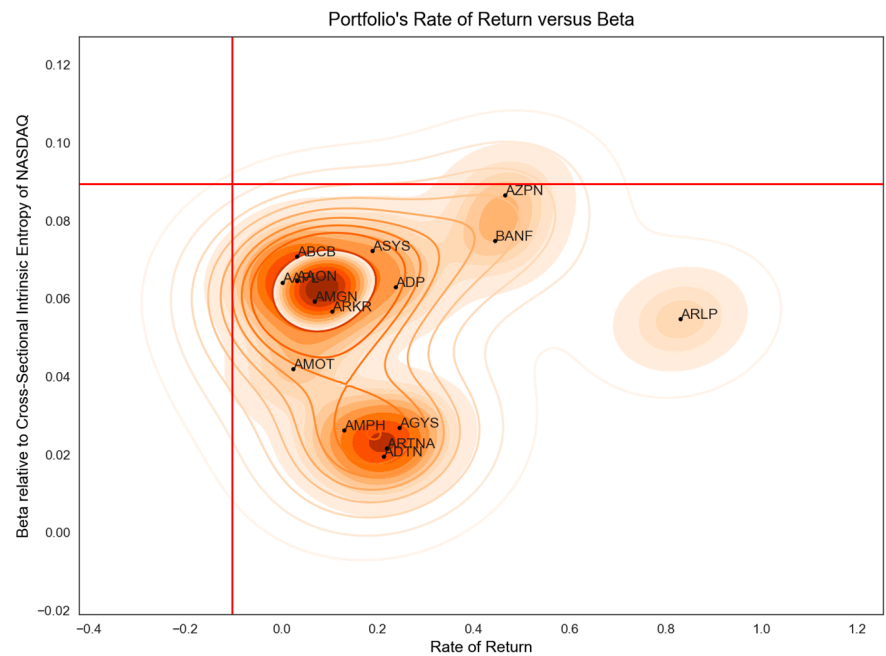

**Figure 4.** A set of 15 NASDAQ symbols that outperformed the NASDAQ Composite index in terms of rate of return without being exposed to higher risk (125-day trading interval, from 1 March 2022 to 26 August 2022).

Figure 5 shows the 15 least risky stock symbols discovered in an interval of 250 trading days, from 5 October 2020 to 20 December 2021, to have a rate of return higher than the NASDAQ Composite index (17.30%) and a beta, relative to the entire NASDAQ market, lower than the one exhibited by the NASDAQ Composite index (0.0319). In total, 228 stocks were identified that meet the constraints and have a positive beta. A rolling window of 20 days has been used. We point out that while the index was up by 17.30% in the considered time interval, most of the symbols identified in the set based on the imposed restrictions were concentrated around a positive rate of return of well above 30%.

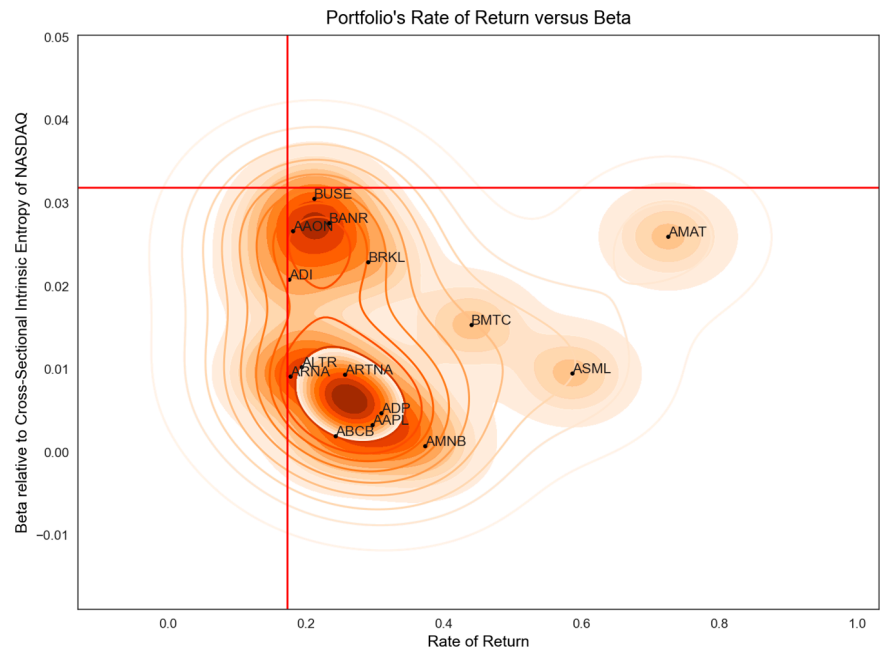

**Figure 5.** A set of 15 NASDAQ symbols that outperformed the NASDAQ Composite index in terms of rate of return without being exposed to higher risk (250-day trading interval, from 5 October 2020 to 20 December 2021).

Figure 6 shows the 15 least risky stock symbols discovered in an interval of 950 trading days, from 2 April 2018 to 26 August 2022, to have a rate of return higher than the NASDAQ Composite index (67.52%) and a beta, relative to the entire NYSE market, lower than the one exhibited by the NASDAQ Composite index (0.0456). In total, 104 stocks were identified that meet the constraints and have a positive beta. A rolling window of 20 days has been used. It should be noted that, while the index increased by 57.52% in the considered time interval, most of the symbols identified in the set based on the imposed restrictions were concentrated around a positive rate of return well above 110%.

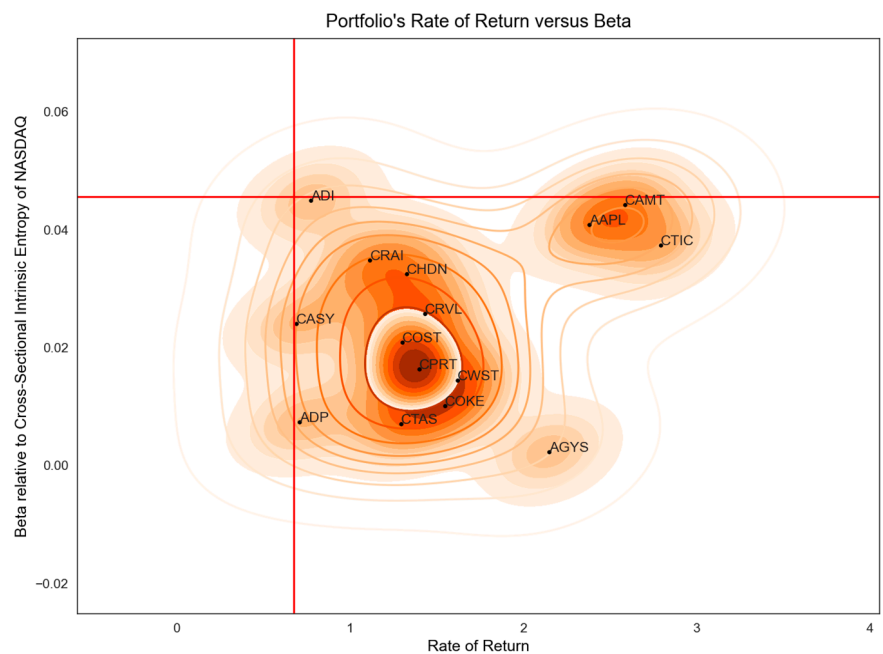

**Figure 6.** A set of 15 NASDAQ symbols that outperformed the NASDAQ Composite index in terms of rate of return without being exposed to a higher risk NYSE S&P500 (950-day trading interval, from 2 April 2018 to 26 August 2022).

In Appendix A, Figures A1–A6 show the portfolio of stocks from the NYSE and the NASDAQ market discovered using as benchmarks for the performance recorded in the studied periods for the DJIA index and Russell 2000 index, respectively.

It should be noted that in the results previously presented, we only considered the stocks with a positive beta. There are two reasons for this:

- First, since the beta of both the market indices S&P500 and NASDAQ Composite were positive in the periods, we wanted to take into account, for finding an answer to our research questions, the stocks that follow the trend of the market, exhibit a lower beta than the beta of the corresponding index, but still have a positive beta.
- Second, the number of stocks with a lower beta than the beta of the corresponding index, including stocks that had a negative beta, is even higher, and the graphical representation would lack clarity.

We point out that the stocks with the highest return rates in the investigated interval of times were those with a negative beta, thus those that were inverse-correlated with the market.

To further study the possibility of diversification from market indices, we test the strategy of systematically buying and holding for one calendar year, for 21 years, between 2001 and 2021. It is a simple strategy that offers equidistant periods. The intervals are equally spaced and able to capture reasonable self-explained phenomena, in terms of the economic forces at work. Within each calendar year, the performance of the S&P500 and NASDAQ Composite indices—the rate of return, and beta relative to the *CSIE* of the corresponding NYSE and the NASDAQ—are computed and further used as references for selecting the stocks that meet the constraints of having a lower or equal risk of the index and a return rate higher or at least equal to that provided by the market index.

The synthetic results obtained for the NYSE market are organised in Table 1 and those for the NASDAQ in Table 2.

**Table 1.** Sets of NYSE stocks, selected annually, based on the rate of return (RoR) of the S&P500 index and its beta relative to the *CSIE* of the market as a whole.

| Year | S&P500 RoR (%) | S&P500 Beta | No. of Symbols in Set | No. of Symbols with Positive Beta | Max. RoR in Symbol Set (%) | Beta of Max. RoR | Min. RoR in Symbol Set (%) | Beta of Min. RoR | Avg. RoR of Symbol Set (%) | Beta of Symbol Set |
|---|---|---|---|---|---|---|---|---|---|---|
| 2001 | −4.14 | 0.038 | 399 | 130 | 250.00 | −0.129 | 0.09 | −0.050 | 22.62 | −0.028 |
| 2002 | −24.82 | 0.042 | 219 | 87 | 348.04 | −0.077 | 0.08 | −0.015 | 13.27 | −0.014 |
| 2003 | 27.22 | 0.025 | 427 | 101 | 603.80 | −0.463 | 27.31 | −0.050 | 70.40 | −0.048 |
| 2004 | 7.98 | 0.028 | 288 | 113 | 162.61 | 0.024 | 8.05 | 0.024 | 30.07 | −0.020 |
| 2005 | 4.93 | 0.026 | 302 | 113 | 171.24 | −0.193 | 5.00 | −0.069 | 26.82 | −0.023 |
| 2006 | 8.49 | 0.021 | 280 | 76 | 132.28 | −0.465 | 8.52 | −0.039 | 24.79 | −0.032 |
| 2007 | 4.73 | 0.036 | 147 | 73 | 228.19 | −0.893 | 4.97 | −0.040 | 36.16 | −0.021 |
| 2008 | −30.43 | 0.026 | 78 | 43 | 250.94 | −0.067 | 0.16 | 0.010 | 22.22 | −0.001 |
| 2009 | 42.22 | 0.008 | 431 | 71 | 2206.38 | −0.211 | 42.26 | −0.009 | 107.89 | −0.034 |
| 2010 | 8.43 | 0.028 | 376 | 167 | 422.53 | −0.121 | 8.43 | −0.089 | 31.87 | −0.014 |
| 2011 | −1.69 | 0.021 | 205 | 128 | 91.59 | −0.017 | 0.16 | −0.003 | 12.81 | 0.001 |
| 2012 | 1.68 | 0.040 | 233 | 131 | 116.43 | 0.019 | 1.72 | 0.029 | 13.26 | −0.007 |
| 2013 | 19.38 | 0.037 | 93 | 44 | 410.45 | −0.213 | 19.42 | −0.046 | 54.48 | −0.034 |
| 2014 | 10.65 | 0.016 | 263 | 69 | 116.13 | −1.331 | 10.67 | −0.033 | 23.43 | −0.031 |
| 2015 | −2.17 | 0.031 | 222 | 149 | 223.81 | −1.585 | 0.08 | 0.008 | 11.75 | −0.011 |
| 2016 | 9.23 | 0.023 | 313 | 99 | 218.33 | −0.347 | 9.27 | −0.083 | 29.71 | −0.035 |
| 2017 | 12.42 | −0.005 | 446 | 0 | 2659.00 | −16.991 | 12.45 | −0.049 | 43.78 | −0.216 |
| 2018 | −8.91 | 0.007 | 268 | 51 | 107.14 | −0.075 | 0.04 | −0.018 | 13.21 | −0.022 |
| 2019 | 14.06 | 0.044 | 316 | 131 | 180.33 | −0.245 | 14.10 | −0.001 | 30.89 | −0.030 |
| 2020 | 55.89 | 0.007 | 540 | 40 | 12,840.00 | −10.986 | 55.90 | −0.017 | 152.73 | −0.164 |
| 2021 | 21.80 | −0.003 | 202 | 0 | 329.77 | −0.165 | 21.86 | −0.024 | 52.02 | −0.086 |

Table 1 presents the sets of NYSE stocks, selected annually, based on the rate of return of the S&P500 index and its beta relative to the *CSIE* of the market as a whole. Table 2 presents the sets of NASDAQ stocks, selected annually, based on the rate of return of the NASDAQ Composite index and its beta relative to the *CSIE* of the market as a whole.

**Table 2.** Sets of NASDAQ stocks, selected annually, based on the rate of return (RoR) of the NASDAQ Composite index and its beta relative to the *CSIE* of the market as a whole.

| Year | NASDAQ Composite RoR (%) | NASDAQ Composite Beta | No. of Symbol in Set | No. of Symbols with Positive Beta | Max. RoR in Symbol Set (%) | Beta of Max. RoR | Min. RoR in Symbol Set (%) | Beta of Max. RoR | Avg. RoR of Symbol Set (%) | Beta of Symbol Set |
|---|---|---|---|---|---|---|---|---|---|---|
| 2001 | −3.20 | 0.067 | 250 | 147 | 370.71 | −0.059 | 0.63 | 0.003 | 45.34 | −0.013 |
| 2002 | −29.00 | 0.049 | 145 | 73 | 324.24 | −0.236 | 0.03 | −0.036 | 27.38 | −0.021 |
| 2003 | 43.40 | −0.006 | 187 | 0 | 1759.80 | −13.005 | 43.66 | −0.010 | 187.91 | −0.168 |
| 2004 | 10.85 | 0.048 | 188 | 86 | 1072.36 | −0.008 | 11.52 | 0.043 | 51.40 | −0.030 |
| 2005 | 9.84 | 0.011 | 140 | 13 | 227.15 | −0.178 | 10.59 | −0.271 | 45.74 | −0.100 |
| 2006 | 4.72 | 0.063 | 185 | 51 | 373.70 | −0.883 | 4.86 | 0.018 | 38.78 | −0.150 |
| 2007 | 10.77 | 0.901 | 144 | 49 | 1900.00 | −123.720 | 10.83 | −1.150 | 68.76 | −2.260 |
| 2008 | −28.64 | 0.447 | 128 | 49 | 296.03 | −1.970 | 0.26 | 0.000 | 26.26 | −0.220 |
| 2009 | 52.96 | 0.065 | 206 | 6 | 3839.29 | −1.630 | 53.19 | −0.188 | 198.60 | −2.549 |
| 2010 | 11.73 | 0.361 | 303 | 171 | 1015.22 | −10.114 | 11.80 | −0.614 | 51.38 | −0.210 |
| 2011 | −1.46 | 0.203 | 235 | 184 | 473.97 | −1.423 | 0.20 | 0.150 | 26.82 | −0.032 |
| 2012 | −1.21 | 0.024 | 328 | 45 | 532.73 | −1.258 | 0.16 | 0.004 | 25.81 | −0.113 |
| 2013 | 29.34 | 0.074 | 288 | 75 | 777.06 | −0.214 | 29.40 | −0.003 | 69.94 | −0.146 |
| 2014 | 9.95 | 0.005 | 164 | 5 | 258.62 | −0.097 | 10.07 | −0.020 | 35.54 | −0.136 |
| 2015 | 0.30 | 0.035 | 320 | 135 | 496.36 | −0.598 | 0.31 | 0.011 | 24.74 | −0.044 |
| 2016 | 12.25 | 0.017 | 290 | 103 | 1628.06 | −2.526 | 12.63 | 0.016 | 50.35 | −0.043 |
| 2017 | 16.99 | 0.008 | 361 | 32 | 483.16 | −0.001 | 17.22 | −0.031 | 60.41 | −0.076 |
| 2018 | −11.32 | 0.008 | 248 | 43 | 839.85 | −0.040 | 0.03 | 0.000 | 29.87 | −0.039 |
| 2019 | 16.17 | 0.022 | 277 | 73 | 526.05 | −0.159 | 16.20 | −0.033 | 47.36 | −0.055 |
| 2020 | 80.24 | 0.007 | 396 | 41 | 6249.56 | −0.298 | 80.33 | −0.153 | 242.09 | −0.084 |
| 2021 | 18.39 | 0.011 | 341 | 93 | 482.67 | −0.260 | 18.41 | −0.010 | 49.91 | −0.029 |

We point out that to extract the best performers from the market, we also considered for portfolio selection the stocks that, in the process of exhibiting a lower risk than the market index, had a negative beta. Additionally, the number of stocks with positive and positive beta is emphasised for each period.

The results presented in Tables 1 and 2 show that the stock market is reliably resourceful in ensuring portfolio diversification. The best performers, in terms of rate of return (RoR), show to have consistently had a negative beta. Even the portfolio beta is consistently negative for each year in the study period of 21 years. This signals the fact that those stocks that performed better than the market index, in terms of return rate, had an inverse correlation with the market as a whole with respect to volatility.

## 4. Discussion

It has to be observed that, except for the year 2003 for the NASDAQ (end of, and recovery after the dot-com bubble burst), and the years 2017 and 2021 for the NYSE, the corresponding market index beta relative to CSIE was consistently positive. This comes as confirmation that in general, the S&P500 and NASDAQ Composite indices are representative of their corresponding markets.

On the other hand, the RoR provided by the best-performing stocks or even by the selected portfolio in its entirety, along with a negative beta exhibited by the majority of the portfolio constituents, support the hypothesis that higher returns can be obtained by investing in stocks that do not follow the market trend in terms of volatility.

Finding that the number of stocks having beta lower than the market index, relative to the CSIE market volatility estimates, and in the negative territory, may potentially suggest a higher risk proposition for the portfolio selection. Thus, in order to answer our research question in comparable terms, we adjusted the selection algorithm by imposing the stock beta to be strictly positive as well:

$$(\beta_{t,w}^{x_j} \leq \beta_{t,w}^{index}) \wedge (\beta_{t,w}^{x_j} > 0) \wedge (R_{x_j} \geq R_{index}), \quad (25)$$

for $j \in [1, m]$; the beta of the stock $x_j$, relative to the entire market volatility. The values $\{m_1, m_2, m_3, \ldots, m_i, \ldots, m_t\}$ are the number of the symbols listed and traded on the market on the corresponding day $i$, and $m = \max(m_i), \forall i \in t$-day time interval.

With this additional constraint, the number of constituent stocks in the selected portfolio is considerably reduced (see the values in the table columns No. of symbols with positive beta). Table 3 presents the sets of NYSE symbols, selected annually, based on the rate of return of the S&P500 index and its beta relative to the *CSIE* of the market as a whole. Only sets of stocks with positive beta relative to the market *CSIE*.

**Table 3.** Sets of NYSE symbols, selected annually, based on the rate of return (RoR) of the S&P500 index and its beta relative to the *CSIE* of the market as a whole. Only sets of stocks with positive beta relative to the market *CSIE*.

| Year | S&P500 RoR (%) | S&P500 Beta | No. of Symbols with Positive Beta | Max. RoR in Symbol Set (%) | Beta of Max. RoR | Min. RoR in Symbol Set (%) | Beta of Min. RoR | Avg. RoR of Symbol Set (%) | Beta of Symbol Set |
|---|---|---|---|---|---|---|---|---|---|
| 2001 | −4.14 | 0.038 | 130 | 196.69 | 0.005 | 0.15 | 0.011 | 21.49 | 0.018 |
| 2002 | −24.82 | 0.042 | 87 | 56.37 | 0.015 | 0.17 | 0.015 | 9.64 | 0.019 |
| 2003 | 27.22 | 0.025 | 101 | 207.3 | 0.019 | 27.57 | 0.021 | 58.77 | 0.013 |
| 2004 | 7.98 | 0.028 | 113 | 162.61 | 0.024 | 8.05 | 0.024 | 27.16 | 0.015 |
| 2005 | 4.93 | 0.026 | 113 | 152.33 | 0.025 | 5.01 | 0.014 | 22.93 | 0.014 |
| 2006 | 8.49 | 0.021 | 76 | 111.64 | 0.008 | 8.84 | 0.004 | 23.03 | 0.01 |
| 2007 | 4.73 | 0.036 | 73 | 135.88 | 0.009 | 5.1 | 0.028 | 29.73 | 0.019 |
| 2008 | −30.43 | 0.026 | 43 | 197.34 | 0.005 | 0.16 | 0.010 | 21.57 | 0.013 |
| 2009 | 42.22 | 0.008 | 71 | 386.24 | 0.002 | 43.83 | 0.003 | 98.01 | 0.004 |
| 2010 | 8.43 | 0.028 | 167 | 196.51 | 0.025 | 8.52 | 0.021 | 29.43 | 0.016 |
| 2011 | −1.69 | 0.021 | 128 | 59.64 | 0.006 | 0.38 | 0.011 | 12.61 | 0.011 |
| 2012 | 1.68 | 0.040 | 131 | 116.43 | 0.019 | 1.72 | 0.029 | 13.42 | 0.02 |
| 2013 | 19.38 | 0.037 | 44 | 348.15 | 0.017 | 20.16 | 0.031 | 51.18 | 0.019 |
| 2014 | 10.65 | 0.016 | 69 | 47.16 | 0.003 | 10.76 | 0 | 21.91 | 0.008 |
| 2015 | −2.17 | 0.031 | 149 | 53.99 | 0.029 | 0.08 | 0.008 | 8.47 | 0.017 |
| 2016 | 9.23 | 0.023 | 99 | 122.58 | 0.018 | 9.31 | 0.016 | 29.24 | 0.013 |
| 2017 | 12.42 | −0.005 | 0 | - | - | - | - | - | - |
| 2018 | −8.91 | 0.007 | 51 | 33.76 | 0.002 | 0.17 | 0.001 | 10.4 | 0.003 |
| 2019 | 14.06 | 0.044 | 131 | 94.78 | 0.037 | 14.11 | 0.020 | 29.58 | 0.022 |
| 2020 | 55.89 | 0.007 | 40 | 1457.92 | 0.006 | 56.88 | 0.004 | 145.54 | 0.004 |
| 2021 | 21.80 | −0.003 | 0 | - | - | - | - | - | - |

Table 4 presents the sets of NASDAQ symbols, selected annually, based on the rate of return of the NASDAQ Composite index and its beta relative to the *CSIE* of the market as a whole. Only sets of stocks with positive beta relative to the market *CSIE*. For years in which the beta of the market index was negative, the composite constraint concluded the impossibility of finding a stock with a beta lower than the beta of the index, but still positive.

**Table 4.** Sets of NASDAQ symbols, selected annually, based on the rate of return (RoR) of the NASDAQ Composite index and its beta relative to the *CSIE* of the market as a whole. Only sets of stocks with positive beta relative to the market *CSIE*.

| Year | NASDAQ Composite RoR (%) | NASDAQ Composite Beta | No. of Symbols with Positive Beta | Max. RoR in Symbol Set (%) | Beta of Max. RoR | Min. RoR in Symbol Set (%) | Beta of Max. RoR | Avg. RoR of Symbol Set (%) | Beta of Symbol Set |
|---|---|---|---|---|---|---|---|---|---|
| 2001 | −3.20 | 0.067 | 147 | 358.63 | 0.001 | 0.63 | 0.003 | 40.93 | 0.030 |
| 2002 | −29.00 | 0.049 | 73 | 115.98 | 0.030 | 0.08 | 0.026 | 23.48 | 0.022 |
| 2003 | 43.40 | −0.006 | 0 | - | - | - | - | - | - |
| 2004 | 10.85 | 0.048 | 86 | 417.92 | 0.037 | 11.52 | 0.043 | 44.73 | 0.023 |
| 2005 | 9.84 | 0.011 | 13 | 94.67 | 0.006 | 15.9 | 0.003 | 34.45 | 0.006 |
| 2006 | 4.72 | 0.063 | 51 | 174.77 | 0.040 | 4.86 | 0.018 | 25.96 | 0.030 |
| 2007 | 10.77 | 0.901 | 49 | 101.7 | 0.274 | 13.09 | 0.223 | 35.47 | 0.408 |
| 2008 | −28.64 | 0.447 | 49 | 90.43 | 0.117 | 0.47 | 0.269 | 20.3 | 0.225 |
| 2009 | 52.96 | 0.065 | 6 | 654.29 | 0.051 | 55.73 | 0.035 | 203.79 | 0.046 |
| 2010 | 11.73 | 0.361 | 171 | 226.99 | 0.282 | 11.86 | 0.321 | 39.87 | 0.203 |
| 2011 | −1.46 | 0.203 | 184 | 276.82 | 0.130 | 0.2 | 0.150 | 20.4 | 0.119 |
| 2012 | −1.21 | 0.024 | 45 | 84.67 | 0.018 | 0.16 | 0.004 | 18.45 | 0.012 |
| 2013 | 29.34 | 0.074 | 75 | 169.36 | 0.010 | 29.61 | 0.043 | 52.95 | 0.039 |
| 2014 | 9.95 | 0.005 | 5 | 55.51 | 0.001 | 12.56 | 0.004 | 31.03 | 0.002 |
| 2015 | 0.30 | 0.035 | 135 | 97.57 | 0.023 | 0.31 | 0.011 | 19.12 | 0.018 |
| 2016 | 12.25 | 0.017 | 103 | 122.75 | 0.013 | 12.63 | 0.016 | 38.82 | 0.008 |
| 2017 | 16.99 | 0.008 | 32 | 147.08 | 0.001 | 18.59 | 0.004 | 40.66 | 0.003 |
| 2018 | −11.32 | 0.008 | 43 | 54.92 | 0.002 | 0.03 | 0.000 | 13.31 | 0.003 |
| 2019 | 16.17 | 0.022 | 73 | 236.86 | 0.004 | 16.36 | 0.015 | 38.25 | 0.011 |
| 2020 | 80.24 | 0.007 | 41 | 1227.7 | 0.002 | 80.92 | 0.004 | 216.04 | 0.003 |
| 2021 | 18.39 | 0.011 | 93 | 123.01 | 0 | 18.46 | 0.002 | 37.26 | 0.006 |

It should be noted that, even when portfolio selection is restricted to stocks that follow the trend volatility of the index and the market as a whole, there can be identified stocks with a lower relative risk of the market that provides significantly higher returns compared to the market index.

In most years, the average RoR of the identified through the proposed methodology portfolio of stocks is consistently positive and a few times higher than the RoR provided by the market index. We underline the fact that thanks to the constraint construction, all the stocks selected for the portfolio have an RoR higher than or at least equal to that realised by the market index.

We point out that we do not consider in our study elements that would constitute more in-depth information related to the sector in which the companies activate, their broader financial performance, experience in the business, workforce, etc. We perceive this aspect as a potential limitation that would have to be addressed in the process of selecting actual portfolios. Furthermore, provided the relatively high number of stocks that could satisfy the selection criteria, we consider that an additional optimization process would be required.

A volatility-managed portfolio that typically applies volatility-timing strategies to the stock market has been studied (Liu et al. 2019) only to discover that these strategies suffer from look-ahead bias, despite existing evidence on the success of the strategies at the stock level. The results of Liu et al. (2019) show that one cannot easily beat the market by timing the market alone. However, their study was grounded on using variance-based volatility estimates. Volatility estimation based on the intrinsic entropy (*IE*) model for longitudinal data (Vințe et al. 2021) and the cross-sectional intrinsic entropy (*CSIE*) as volatility estimator for the market as a whole ((Vințe and Ausloos 2022), benefit from the additional information provided by traded volume taken into account, with the meaning of traction that the market gives to a certain price level, along with the signed volatility estimates: negative values indicating an inclination on the market to sell, and positive values suggesting a preponderantly buy tendency on the market.

Additionally, it should be observed that the denominator of these betas represents the variance of the volatility estimates of the entire market $V_{CSIE,t,w}^{market}$ based on the cross-sectional intrinsic entropy (*CSIE*) and therefore volatility of volatility (VoV). Volatility is a fundamental quantity that describes the dynamics of volatility processes. However, Li et al. (2022) argue that it is far less well understood and constructed a nonparametric estimator of the VoV based on noisy high-frequency data with price jumps. The perspective provided to intraday trading by the high-frequency data for dynamically estimating portfolio volatility through *CSIE* and its volatility could represent a pertinent path to follow for further research.

## 5. Conclusions

In the context of the cross-sectional intrinsic entropy model (CSIE) to estimate the volatility of the stock market as a whole, we consider EOD data containing daily OHLC prices along with the traded quantity (volume) of each marked listed symbol that may be selected in a set as a portfolio constituent.

The research presented in this article makes use of historical daily open, high, low, and close prices and volume for a period of over 21 years, from 1 January 2001 to 28 October 2022.

In our study, we benchmark portfolio volatility risks against the volatility of the entire market provided by the *CSIE* and the volatility of market indices computed using longitudinal data. We introduce *CSIE*-based betas to characterise the relative volatility risk of the portfolio against market indices and the market as a whole.

The results we obtained empirically answer the research questions we established.

i. For any given interval of time, at least two symbols, traded on the market, that have a combined risk equal to or lower than that provided by the volatility estimates of the market index and with a higher rate of return, can be identified.
ii. Algorithmically, we discover all symbols that satisfy these constraints.

Thus, we empirically prove that, through *CSIE*-based betas, multiple sets of symbols that outperform the market indices in terms of rate of return while maintaining the same level of risk or even lower than the one exhibited by the market index can be discovered, for any given time interval.

**Author Contributions:** Conceptualization, C.V.; Data curation, C.V.; Formal analysis, C.V. and M.A.; Investigation, C.V.; Methodology, C.V. and M.A.; Software, C.V.; Supervision, M.A.; Writing—original draft, C.V.; Writing—review & editing, C.V. and M.A. All authors have read and agreed to the published version of the manuscript.

**Funding:** This research received no external funding.

**Data Availability Statement:** The data presented in this study is available on request from the corresponding author. The data is not publicly available due to the original software components necessary to obtain them and developed for ongoing research purposes.

**Acknowledgments:** The authors thank and extend their gratitude to the three anonymous reviewers for their careful reading of the manuscript and their insightful comments and suggestions, which no doubt greatly help to improve the report!

**Conflicts of Interest:** The authors declare no conflict of interest.

## Appendix A

Figure A1 shows the 15 least risky stock symbols discovered in an interval of 125 trading days, from 1 March 2022 to 26 August 2022, which have a return rate higher than the NYSE DJIA index (−3.04%) and a beta, relative to the entire NYSE market, lower than the one exhibited by the DJIA index (0.0459). In total, 220 stocks were identified that satisfy the constraints and have a positive beta. In other words, a set of symbols can be identified, more than one, that have a higher rate of return than the DJIA market index at a maximum level of risk provided as a threshold represented here by the beta of the index relative to the entire NYSE market. It is worth noting that while the index was down by 3.04% in the considered time interval, most of the symbols identified in the set based on the imposed restrictions were concentrated around a positive rate of return of well above 10%. A rolling window of 10 days has been used.

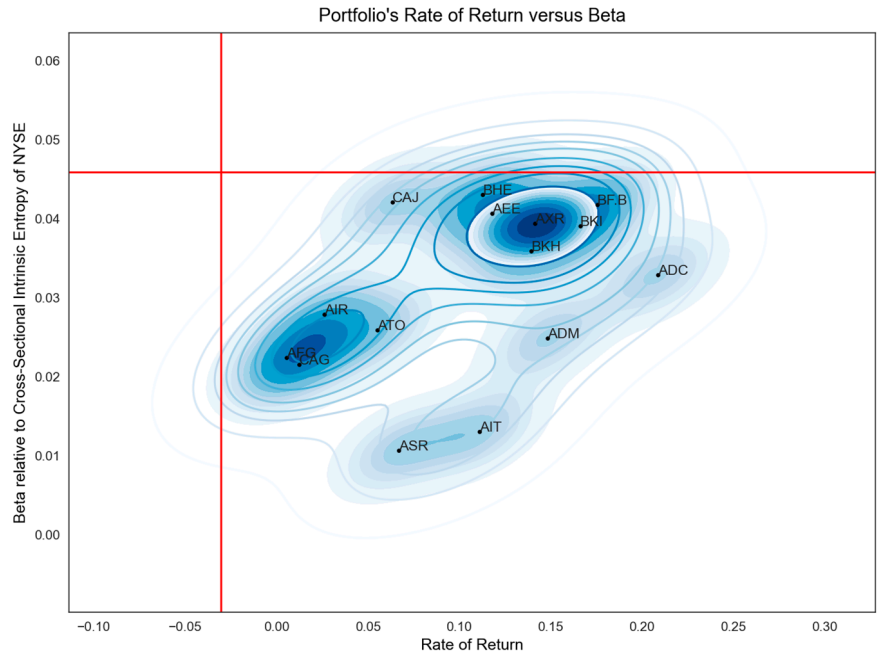

**Figure A1.** A set of 15 NYSE symbols that outperformed the DJIA index in terms of rate of return without being exposed to higher risk (125-day trading interval, from 1 March 2022 to 26 August 2022).

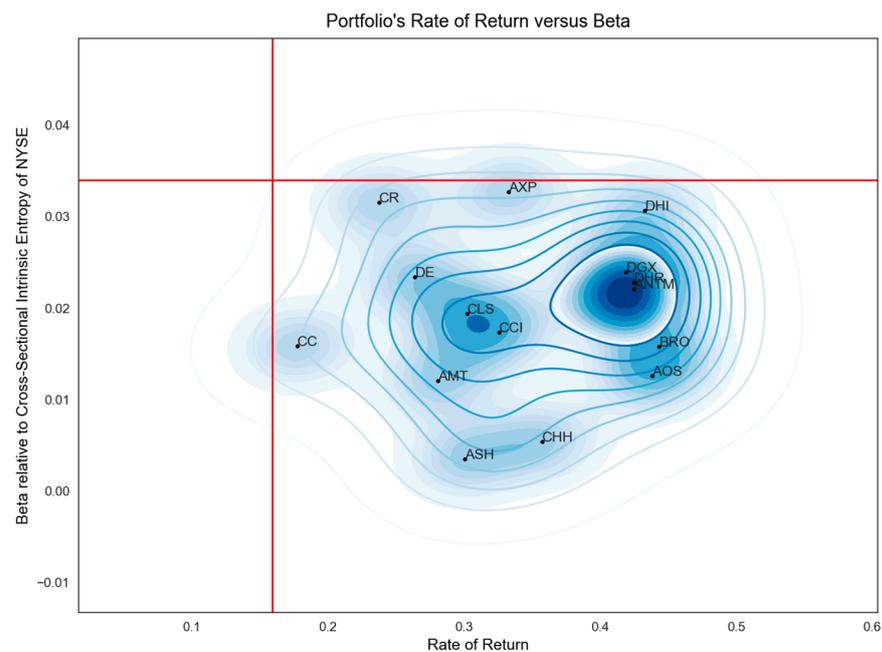

**Figure A2.** A set of 15 NYSE symbols that outperformed the DJIA index in terms of rate-of-return without being exposed to higher risk (250-day trading interval, from 5 October 2020 to 20 December 2021).

Figure A2 shows the 15 least risky stock symbols discovered in an interval of 250 trading days, from 5 October 2020 to 20 December 2021, which have a rate of return higher than the NYSE DJIA index (15.94%) and a beta, relative to the entire NYSE market, lower than the one exhibited by the DJIA index (0.0339).

In total, 151 stocks were identified that satisfy the constraints and have a positive beta. A rolling window of 20 days has been used. We point out that, while the index was up by 15.94% in the considered time interval, most of the symbols identified in the set based on the imposed restrictions were concentrated around a positive rate of return of well above 30%.

Figure A3 shows the 15 least risky stock symbols discovered in an interval of 950 trading days, from 2 April 2018 to 26 August 2022, which have a rate of return higher than the NYSE DJIA index (27.03%) and a beta, relative to the entire NYSE market, lower than the one exhibited by the DJIA index (0.0298). In total, 136 stocks were identified that meet the constraints and have a positive beta. A rolling window of 20 days has been used. We point out that, while the index was up by 27.03% in the considered time interval, most of the symbols identified in the set based on the imposed restrictions were concentrated around a positive rate of return of well above 50%.

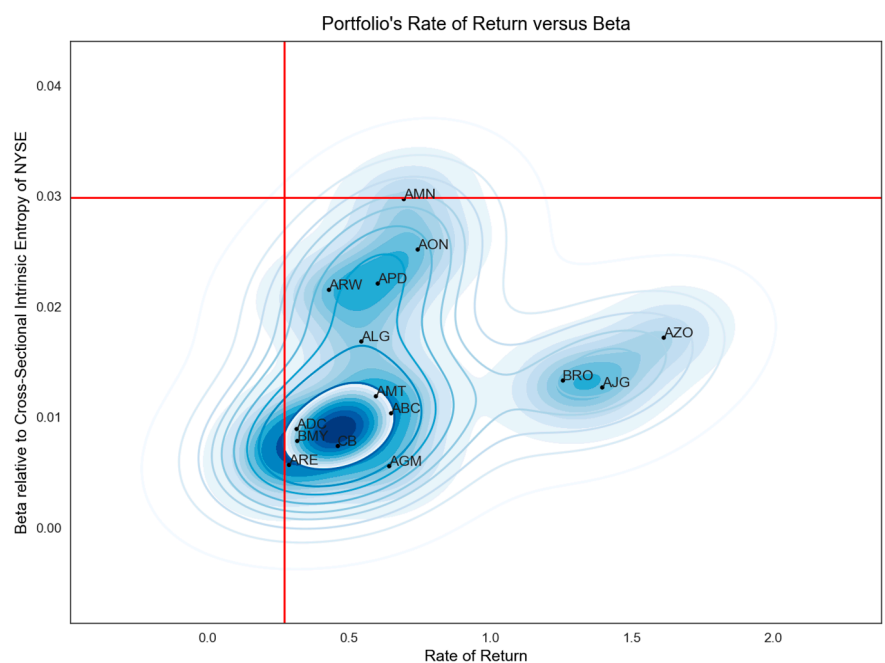

**Figure A3.** A set of 15 NYSE symbols that outperformed the DJIA index in terms of rate-of-return without being exposed to higher risk (950-day trading interval, from 2 April 2018 to 26 August 2022).

Figure A4 shows the 15 least risky stock symbols discovered in an interval of 125 trading days, from 1 March 2022 to 26 August 2022, to have a rate of return higher than the NASDAQ Russell 2000 index (−10.28%) and a beta, relative to the entire NASDAQ market, lower than the one exhibited by the Russell 2000 index (0.0894). In total, 109 stocks were identified that satisfy the constraints and have a positive beta. In other words, a set of symbols can be identified, more than one, that have a higher rate of return than the Russell 2000 market index at a maximum level of risk provided as a threshold represented here by the beta of the index relative to the entire NASDAQ market. It is worth noting that while the index was down by 10.28% in the considered time interval, most of the symbols identified in the set based on the imposed restrictions were concentrated around a positive rate of return of about 11%. A rolling window of 10 days has been used.

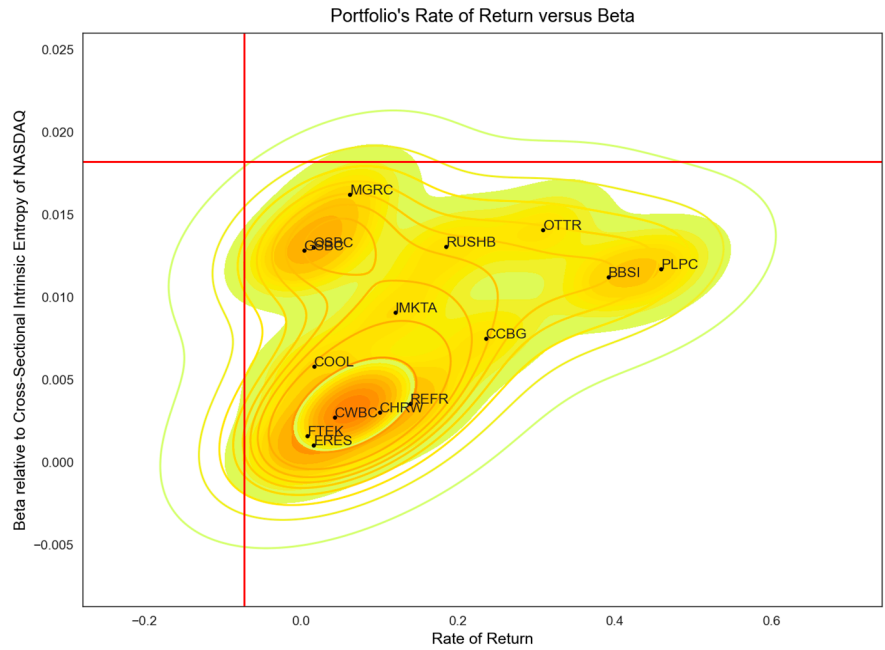

**Figure A4.** A set of 15 NASDAQ symbols that outperformed the Russell 2000 Composite index in terms of rate of return without being exposed to higher risk (125-day trading interval, from 1 March 2022 to 26 August 2022).

Figure A5 shows the 15 least risky stock symbols discovered in an interval of 250 trading days, from 5 October 2020 to 20 December 2021, that have a rate of return higher than the NASDAQ Russell 2000 index (8.60%) and a beta, relative to the entire NASDAQ market, lower than the one exhibited by the Russell 2000 index (0.0118).

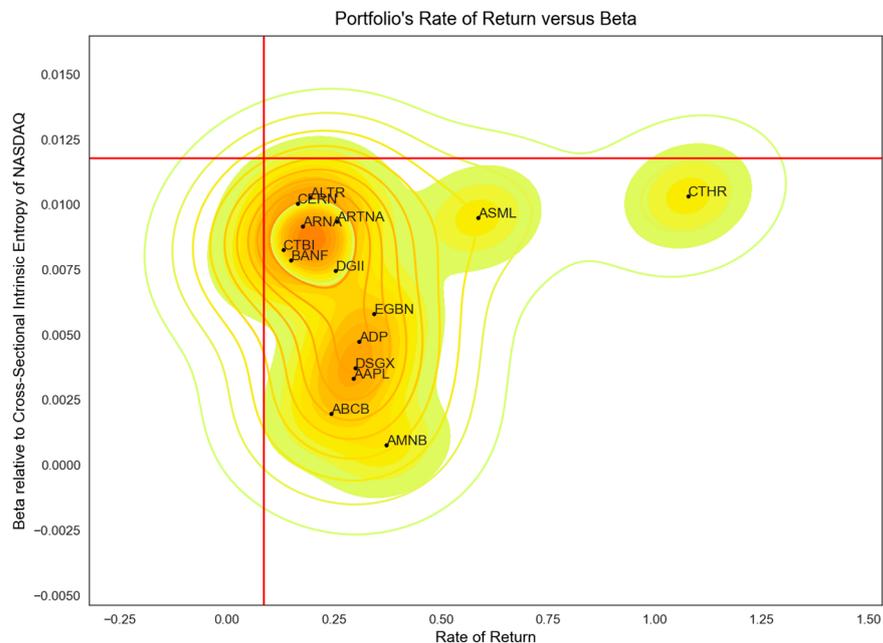

**Figure A5.** A set of 15 NASDAQ symbols that outperformed the Russell 2000 index in terms of rate of return without being exposed to higher risk (250-day trading interval, from 5 October 2020 to 20 December 2021).

In total, 124 stocks were identified that satisfy the constraints and have a positive beta. A rolling window of 20 days has been used. We point out that while the index was up by 8.60% in the considered time interval, most of the symbols identified in the set based on the imposed restrictions were concentrated around a positive rate of return of above 35%.

Figure A6 shows the only 15 symbols discovered in an interval of 950 trading days, from 2 April 2018 to 26 August 2022, to have a rate of return higher than the NASDAQ Russell 2000 index (25.42%) and a beta, relative to the entire NASDAQ market, lower than the one exhibited by the Russell 2000 index (0.0051). A rolling window of 20 days has been used. We point out that while the index was up by 25.42% in the considered time interval, most of the symbols identified in the set based on the imposed restrictions were concentrated around a positive rate of return of well above 65%.

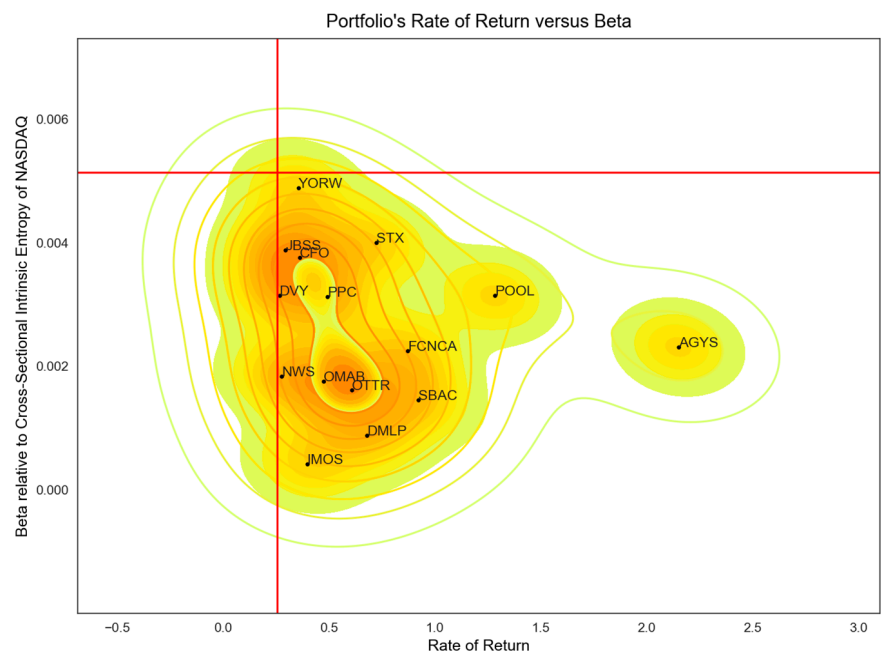

**Figure A6.** A set of 15 NASDAQ symbols that outperformed the Russell 2000 index in terms of rate of return without being exposed to higher risk (950-day trading interval, from 2 April 2018 to 26 August 2022).